\documentclass{aa}
\usepackage{amsmath}
\usepackage[T1]{fontenc}
\usepackage{epsfig}
\usepackage{epsfig}
\usepackage{natbib}
\usepackage{times}

\bibpunct{(}{)}{;}{a}{}{,}

\begin{document}

   \thesaurus{ 12        % A&A Section Physical and chemical processes
              (02.08.1;  % Physical data and processes: Hydrodynamics
               02.14.1;  % Physical data and processes: Nuclear reactions, 
                         % nucleosynthesis, abundances
               02.20.1;  % Physical data and processes: Turbulence
               03.13.4;  % Methods:numerical
               08.19.4)  % Supernovae: general
              }

   \title{A new model for deflagration fronts in reactive fluids}

   \author{M.\ Reinecke\inst{1}\and
           W.\ Hillebrandt\inst{1}\and
           J.C.\ Niemeyer\inst{2}\and
           R.\ Klein\inst{3}\and
           A.\ Gr\"obl\inst{1}
          }

   \offprints{M.\ Reinecke}

   \institute{Max-Planck-Institut f\"ur Astrophysik,
              Karl-Schwarzschild-Str. 1, 85740 Garching, Germany\
              \and
              LASR, Enrico Fermi Institute, University of Chicago, 
              Chicago, IL 60637, USA\
              \and
              Konrad-Zuse-Zentrum f\"ur Informationstechnik Berlin,
              Takustra\ss e 7, 14195 Berlin--Dahlem, Germany
             }

   \date{Received November dd, 1998; accepted mm dd, yyyy}

   \maketitle

   \begin{abstract}
      We present a new way of modeling deflagration fronts in reactive fluids,
      the main emphasis being on turbulent thermonuclear
      deflagration fronts in white dwarfs undergoing a Type Ia
      supernova explosion. Our approach is based on a level set method
      which treats the front as a mathematical discontinuity and
      allows full coupling between the front geometry and the flow
      field \citep {smiljanovski-etal-97}. With only minor modifications,
      this method can also be applied to describe contact discontinuities.
      Two different implementations are described and
      their physically correct behaviour for simple testcases is shown.
      First results of the method applied to the concrete problems
      of Type Ia supernovae and chemical hydrogen combustion are briefly
      discussed;
      a more extensive analysis of our astrophysical simulations is given in
      \cite{reinecke-etal-98b}.

   \end{abstract}

\section{Introduction}
Numerical simulations of turbulent combustion have always been a challenge,
mainly because of the large range of length scales involved. In astrophysics,
prominent examples are Type Ia supernovae, where the length scales of relevant
physical processes range from 10$^{-4}$cm to several 10${^8}$cm).
In the currently favoured scenario the explosion starts as a deflagration
in the flamelet regime near the center of the star. At the corresponding
densities, the typical width of the conductive flame is less than 1mm
\citep{timmes-woosley-92}. Rayleigh-Taylor unstable blobs of hot burnt
material are thought to form which rise and lead to shear-induced turbulence
at their interface with the unburnt gas. This turbulence increases the
effective surface area of the flamelets and thereby the rate of fuel
consumption over its laminar value; the hope is that finally a fast
deflagration might result, in agreement with phenomenological models of Type
Ia explosions \citep{nomoto-etal-84}.

A multidimensional direct numerical simulation of such an event
is --~and will always be~-- computationally infeasible; therefore,
small scale effects like turbulence, diffusion and heat conduction
need to be incorporated in form of phenomenological models.
Despite considerable progress in the field of modeling turbulent
combustion for astrophysical flows (see, e.g., \citealt{niemeyer-95}),
the correct numerical representation of the thermonuclear deflagration front
is still a weakness of Type Ia simulations;
this is mainly due to the fact that in those simulations the conductive flame
is not properly resolved, but must be made several orders
of magnitude thicker than in reality. 
The artificially increased width of the reaction zone is a prerequisite
for the reactive-diffusive flame model \citep{khokhlov-93}, which has been used
by most authors so far. In this approach the burning region is
stretched out over several grid zones to ensure an isotropic
flame propagation speed. Typical values for the numerical flame width range
from 4--5 \citep{khokhlov-93} to 8--10 grid cells \citep{niemeyer-94}.
However, the artificially soft transition from fuel to ashes stabilizes
the front against hydrodynamical instabilities on small length scales,
which in turn results in an underestimation of the flame surface area and
--~consequently~-- of the total energy generation rate.

The front tracking method described in this paper is based on the so-called
\emph{level set technique} that has been in use for several years in the
engineering sciences. It was introduced by \cite{osher-sethian-88} who used
the zero level set of a $n$-dimensional scalar function to represent
$(n-1)$-dimensional front geometries. \cite{sussman-etal-94} give equations
for the time evolution of such a level set which is passively advected by a
flow field; this approach can be used to track contact discontinuities,
for example.
\cite{smiljanovski-etal-97} extend this method to allow the tracking
of fronts additionally propagating normal to themselves, e.g. deflagrations
and detonations. 
In contrast to the
artificial broadening of the flame in the reaction-diffusion-approach, their
algorithm is able to treat the front as an exact hydrodynamical discontinuity.
Considering the fact that the real width of the conductive flame in a Type Ia
supernova is several orders of magnitude smaller
than the typical grid cell sizes in multidimensional simulations, this
is a very good approximation.

The outline of this paper is as follows: In section \ref{levdescr} we present
the main ideas and governing equations of our approach. Two different
implementations of the flame model are described in detail in section
\ref{impl}. Section \ref{test} is dedicated to the results of simple
testcases, whereas section \ref{appl} lists some results of the application
of our numerical scheme to ``real world'' problems.
Finally, we give a summary of open issues and
an outlook on future work in section \ref{summ}.

\section{The level set method}
\label{levdescr}
The central aspect of our front tracking method is the association
of the front geometry (a time-dependent set of points $\Gamma$)
with an isoline of a so-called level set function $G$:
\begin{equation}
  \Gamma:=\lbrace\vec r\ |\ G(\vec{r}) = 0\rbrace
\end{equation}
Since $G$ is not completely determined by this equation, we can additionally
postulate that $G$ be negative in the unburnt and positive in the burnt
regions, and that $G$ be a ``smooth'' function, which is convenient from
a numerical point of view. This smoothness can be achieved, for example,
by the additional constraint that
\begin{equation}
  |\vec{\nabla}G| \equiv 1
  \label{absgrad}
\end{equation}
in the whole computational domain, with the exception of possible extrema
and kinks of
$G$. The ensemble of these conditions produces a $G$ which is a signed distance
function, i.e. the absolute value of $G$ at any point equals the minimal
front distance.

The normal vector to the front is defined as
\begin{equation}
 \vec n := -\frac{\vec \nabla G}{|\vec\nabla G|}
 \label{normal}
\end{equation}
and thus points towards the unburnt material.

The task is now to find an equation for the temporal evolution of $G$
such that the zero level set of $G$ behaves exactly
as the flame. Such an expression can be obtained by the consideration
that the total velocity of the front consists of two independent contributions:
it is advected by the fluid motions at a speed $\vec v$ and
it propagates normal to itself with a burning speed $s$.

Since for deflagration waves a velocity jump usually
occurs between the pre-front and post-front states, we must explicitly
specify which state $\vec v$ and $s$ refer to; traditionally, the values
for the unburnt state are chosen. Therefore, one obtains for the total
front motion
\begin{equation}
\label{Df}
  \vec D_f = \vec v_u + s_u \vec n.
\end{equation}
The total temporal derivative of $G$ at a point $\vec P$ attached to the front
must vanish, since $G$ is, by definition, always 0 at the front:
\begin{equation}
  \frac{\text{d}G_{\vec P}}{\text{d}t} = \frac{\partial G}{\partial t}
  + \vec \nabla G \cdot \dot {\vec x}_{\vec P} = \frac{\partial G}{\partial t}
  + \vec D_f\cdot \vec\nabla G = 0
\end{equation}
This leads to the desired differential equation describing the time
evolution of $G$:
\begin{equation}
  \frac{\partial G}{\partial t} = - \vec D_f\cdot \vec\nabla G
  \label{levprop}
\end{equation}

This equation, however, cannot be applied on the whole computational
domain: Firstly, $\vec D_f$ has a physical meaning in the immediate
vicinity of the front only and may be undefined elsewhere.
Secondly, using this equation everywhere will
in most cases destroy $G$'s distance function property (eq. \ref{absgrad}).
As a consequence, this might lead to the buildup of very steep slopes
in $G$ which are likely to cause numerical problems \citep{sussman-etal-94}.
Therefore additional measures must be taken in the regions away from the front
to ensure a ``well-behaved'' $|\vec\nabla G|$ (for implementation details, see
section \ref{passive-reinit}).

The situation is further complicated by the fact that the quantities
$\vec v_u$ and $s_u$ which are needed to determine $\vec D_f$ are not
readily available in the cells cut by the front. In a finite volume context,
these cells contain a mixture of pre- and post-front states instead.
Nevertheless one can assume that the conserved quantities (mass, momentum and
total energy) of the mixed state satisfy the following conditions:
\begin{alignat}{2}
  \overline{\rho}   &= \alpha \rho_u &&+ (1-\alpha) \rho_b \label{cons1}\\
  \overline{\rho \vec v} &= \alpha \rho_u \vec v_u
    &&+ (1-\alpha) \rho_b \vec v_b \label{cons2} \\
  \overline{\rho e} &= \alpha \rho_u e_u &&+ (1-\alpha) \rho_b e_b \label{cons3}
\end{alignat}
Here $\alpha$ denotes the volume fraction of the cell occupied by the unburnt
state.
In order to reconstruct the states before and behind the flame, a nonlinear
system consisting of the equations above, the Rankine-Hugoniot jump conditions
and a burning rate law must be solved. The technical details are described in
section \ref{complete-reconstruct}.

  \begin{figure}
  \centerline{\epsfig{file=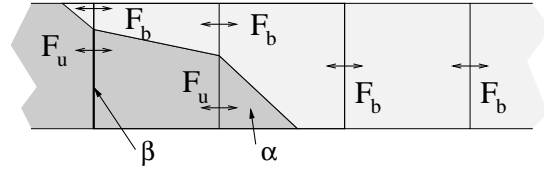,width=0.4\textwidth}}
  \caption{
   Illustration of the basic principles of the level set method according
   to \cite{smiljanovski-etal-97}: The piecewise linear front cuts the mixed
   cells into burnt and unburnt parts. $\alpha$ is the unburnt volume fraction
   of a cell, $\beta$ is the unburnt area fraction of a cell interface. The
   fluxes $\vec F_u$ and $\vec F_b$ are calculated from the reconstructed
   states.}
  \label{flspl}
  \end{figure}

Having obtained the reconstructed pre- and post-front states in the mixed
cells, it is not only possible to determine $\vec D_f$, but also to separately
calculate the fluxes of burnt and unburnt material over the cell interfaces.
Consequently, the total flux over an interface can be expressed as a linear
combination of burnt and unburnt fluxes weighted by the unburnt interface area
fraction
$\beta$:
\begin{equation}
  \label{spliteq}
  \vec{\bar{F}} = \beta \vec F_u + (1-\beta)\vec F_b
\end{equation}
(see Fig. \ref{flspl}).

\section{Implementation}
\label{impl}
In this section we concentrate on the case of a deflagration wave,
but the modifications needed to model contact discontinuities are
straightforward: for this case, the front propagation speed ($s$ or $s_u$)
and the formation enthalpy ($q$) have to be zero in all following equations,
which leads to an overall simplification of the numerical scheme.

For our calculations, the front tracking algorithm was implemented as
an additional module for the hydrodynamics code PROMETHEUS
\citep{fryxell-etal-89}. Two independent and completely different
implementations were realized:
\begin{itemize}
  \item In the simpler approach, the $G$-function plays a somehow passive role:
    It is advected by the fluid motions and by burning and is only used to
    determine the source terms for the reactive Euler equations. We will
    refer to this algorithm as \emph{passive implementation}.
    It must be noted that there exists no \emph{real} discontinuity between
    fuel and ashes in this case; the transition is smeared out over
    about three grid cells by the hydrodynamical scheme, and the level set
    only indicates where the thin flame front \emph{should} be. However,
    the numerical flame is still considerably thinner than in the
    reaction-diffusion approach.
  \item The second implementation (called \emph{complete implementation})
    contains in-cell-reconstruction and
    flux-splitting as proposed by \cite{smiljanovski-etal-97}; therefore it
    should exactly describe the coupling between the flame and the hydrodynamic
    flow.
\end{itemize}

\subsection{Passive Implementation}
\label{passive}
\subsubsection{$G$-Transport}
Since the front motion consists of two distinct contributions,
it is appropriate to use an operator splitting approach for the
time evolution of $G$. The advection term due to the fluid velocity
$\vec v_F$ can be written as
\begin{equation}
\frac{\partial G}{\partial t} = -\vec v_F \vec\nabla G,
\end{equation}
or in conservative form
\begin{equation}
\int_V\frac{\partial (\rho G)}{\partial t}d^3 r
+\oint_{\partial V} -\vec v_F \rho G d\vec f = 0
\end{equation}
\citep{mulder-etal-92}. This equation is identical to the advection equation
of a passive scalar, like the concentration of an inert chemical species.
Consequently, this contribution to the front propagation can be calculated
by PROMETHEUS itself without requiring complicated modifications.
As a consequence, the discrete values of the level set function have to be
stored at the centers of the grid cells, like the hydrodynamical variables
$\rho$, $T$, etc.

The additional flame propagation due to burning is calculated at the end
of each time step according to the following procedure:

First the four discrete spatial derivatives of $G$ are obtained in each cell:
    \begin{align}
       D^+_{x,ij}&:=\frac{G_{i+1,j} - G_{i,j}}{x_{i+1}-x_i} \quad
       D^-_{x,ij}:=\frac{G_{i,j} - G_{i-1,j}}{x_i-x_{i-1}}\\
       D^+_{y,ij}&:=\frac{G_{i,j+1} - G_{i,j}}{y_{j+1}-y_j} \quad
       D^-_{y,ij}:=\frac{G_{i,j} - G_{i,j-1}}{y_j-y_{j-1}}
    \end{align}
    At the boundaries of the computational domain some of the above equations
    cannot be applied (e.g. $D^-_{x,1j}$). In these cases, the gradient is set
    to 0 for reflecting boundaries and extrapolated in zeroeth order for
    outflow boundaries.

Afterwards, the relevant derivatives are determined by simple upwinding
with respect to the propagation direction of the front:
         \begin{equation}
           D_{x,ij} = \begin{cases}
           D^+_{x,ij} & \text{for } D^+_{x,ij}>0 \text{ and } D^-_{x,ij}>0 \\
           D^-_{x,ij} & \text{for } D^+_{x,ij}<0 \text{ and } D^-_{x,ij}<0 \\
           \bar D_{x,ij} &
             \text{for } (D^-_{x,ij}\cdot D^+_{x,ij}) \leq 0 \\
           \end{cases}
         \end{equation}
    where $\bar D_{x,ij}:=0.5(|D^-_{x,ij}|+|D^+_{x,ij}|)$.

The new $G$-value is then defined by
        \begin{equation}
          G_{ij}'=G_{ij}+\Delta t s_{ij} \sqrt{D_{x,ij}^2 + D_{y,ij}^2}.
        \end{equation}

\subsubsection{Re-Initialization}
\label{passive-reinit}
As was mentioned in section \ref{levdescr},
an additional correction step has to be applied in the regions away from the
front in order to keep $G$ a signed distance function.
This task can be accomplished
in several ways. \cite{sussman-etal-94}, for example, suggest a pseudo-time
approach, where the equation
\begin{equation}
      \frac{\partial G}{\partial \tau} = \frac{G}{|G|+\varepsilon}
        \left(1-|\vec\nabla G|\right)
\end{equation}
is solved iteratively until convergence is obtained. Here, $\varepsilon$
denotes an empirical quantity with a value comparable to the length
of a grid cell. While being quite
efficient, this method has the drawback that it changes \emph{all} $G$-values,
even those near the front; consequently, the front might be moved by
small amounts during the re-initialization \citep{sethian-96}.

This potential problem is avoided by the following algorithm which we used
for our simulations:
\begin{itemize}
  \item The coordinates of all zero crossings of $G$ between neighbouring grid
  points are calculated by linear interpolation; if, e.g., $G_{i,j}>0$ and
  $G_{i+1,j}<0$, the zero crossing is at
  \begin{align}
     x_z &= x_i+\left|\frac{G_{i,j}}{G_{i+1,j}-G_{i,j}}\right|(x_{i+1}-x_i)
     \quad \text{and}\\
   \quad y_z &= y_j
  \end{align}
  The ensemble of all points ($x_z, y_z$) is a discrete representation of the
  zero level set.

  \item For all grid points, the minimum distance to one of the points
  ($x_z, y_z$) is determined:
  \begin{equation}
    d_{ij}=\text{min}_n\sqrt{(x_i-x_{z,n})^2 + (y_j-y_{z,n})^2}
  \end{equation}

  \item The corrected value for $G_{ij}$ is a weighted average of
  the original value and $d_{ij}$, such that
  \begin{equation}
     G_{ij}' := H(d_{ij})G_{ij} + (1-H(d_{ij}))\text{sgn}(G_{ij})d_{ij}.
  \end{equation}
  $H$ denotes a function, which is essentially 1 for small arguments and
  smoothly drops to 0 near a given threshold. In this work, we used the
  expression
      \begin{equation}
        H(d)=\left(1-\tanh \frac{d-d_0}{\delta/3}\right) \mathbf{\Bigm/}
          \left(1-\tanh\frac{-d_0}{\delta/3}\right).
      \end{equation}
  For this equation, the transition takes place in a region of the width
  $\delta$ around $d_0$. Satisfying results have been obtained for
  $d_0\approx3\Delta$ and $\delta\approx\Delta$, where $\Delta$ represents
  the width of a grid cell.
\end{itemize}

The weighting with $H(d)$ has the effect that values near the flame are
practically left unchanged, while the values farther away represent a distance
function in good approximation.

\subsubsection{Source terms}
After the update of the level set function in each time step, the change
of chemical composition and total energy due to burning is calculated in the
cells cut by the front. In order to obtain these values, the volume fraction
$\alpha$ occupied by the unburnt material is determined in those cells by the
following approach: from the value $G_{ij}$ and the two steepest gradients
of $G$ towards the front in $x$- and $y$-direction a first-order approximation
$\tilde G$ of the level set function is calculated; then the area fraction
of cell $ij$ where $\tilde G < 0$ can be found easily.
Based on these results, the new concentrations of fuel, ashes and energy are
obtained:
\begin{align}
X'_{\text{Ashes}} &= \text{max}(1-\alpha, X_{\text{Ashes}}) \label{newash} \\
X'_{\text{Fuel}} &= 1-X'_{\text{Ashes}} \\
e'_{\text{tot}} &= e_{\text{tot}} + q (X'_{\text{Ashes}}-X_{\text{Ashes}})
\end{align}
In principle this means that all fuel found behind the front is converted
to ashes and the appropriate amount of energy is released. The maximum operator
in eq. (\ref{newash}) ensures that no ``reverse burning'' (i.e. conversion from
ashes to fuel) takes place in the cases
where the average ash concentration is higher than the burnt volume fraction;
such a situation can occur in a few rare cases because of unavoidable
discretization errors of the numerical scheme.

\subsection{Complete Implementation}
\label{complete}
In this approach the discrete values of $G$ are defined on the
cell corners instead of the cell centers, because this simplifies the
calculation of the geometrical quantities $\alpha$ and $\beta$, which are
needed for the reconstruction and flux-splitting steps.
In the following sections all quantities defined on cell corners are described
by fractional indices: e.g. $G_{i+1/2,j+1/2}$ denotes the $G$ value in the
top right corner of cell $ij$.

\subsubsection{Geometrical quantities}
The knowledge of the front normal $\vec n$ and the unburnt volume fraction
$\alpha$ in the mixed cells is a prerequisite for the reconstruction of burnt
and unburnt hydrodynamical states. The normal is derived from the discrete
gradient
    \begin{alignat}{2}
      \left(\frac{\partial G}{\partial x}\right)_{ij} &=
      \frac{1}{2}\bigg(&&
      \frac{G_{i+1/2,j+1/2}-G_{i-1/2,j+1/2}}{x_{i+1/2}-x_{i-1/2}} + \notag\\
   &&&   \frac{G_{i+1/2,j-1/2}-G_{i-1/2,j-1/2}}{x_{i+1/2}-x_{i-1/2}}
      \bigg) \\
      \left(\frac{\partial G}{\partial y}\right)_{ij} &=
      \frac{1}{2}\bigg(&&
      \frac{G_{i+1/2,j+1/2}-G_{i+1/2,j-1/2}}{y_{j+1/2}-y_{j-1/2}} + \notag\\
   &&&   \frac{G_{i-1/2,j+1/2}-G_{i-1/2,j-1/2}}{y_{j+1/2}-y_{j-1/2}}
      \bigg).
    \end{alignat}
According to eq. (\ref{normal}), $\vec n_{ij}$ is then given by
\begin{equation}
\vec n_{ij}=-\frac{(\vec\nabla G)_{ij}}{|(\vec\nabla G)_{ij}|}.
\end{equation}

  \begin{figure}
  \centerline{\epsfig{file=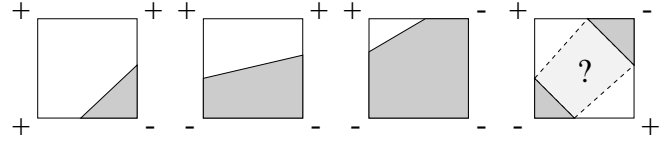,width=0.48\textwidth}}
  \caption{Determination of $\alpha$ in a mixed cell. The signs on the
    cell corners denote the sign of the $G$ function. The rightmost sketch
    shows a situation where two different geometrical interpretations are
    possible and $\alpha$ is not uniquely defined.}
  \label{alpha}
  \end{figure}

The value for $\alpha$ is found by determining the zeros of $G$ on
all cell edges, connecting them with straight lines and calculating the surface
area behind this approximated flame. Fig. \ref{alpha} shows all topologically
different situations.
While calculating $\alpha$ in the first three cases is trivial,
the fourth case is ambiguous since two different front geometries are possible;
for this situation, we set $\alpha$ to the mean value of the two possibilities.
Fortunately, such a geometrical constellation is quite rare in hydrodynamical
simulations.

\subsubsection{Reconstruction}
\label{complete-reconstruct}
In order to obtain the hydrodynamical state vectors $\vec U_u$ and $\vec U_b$
from the average $\bar {\vec U}$ in the mixed cells, a nonlinear equation system
has to be solved. The first three equations have already been presented in
section \ref{levdescr} (eqs. \ref{cons1} -- \ref{cons3}). It is convenient to
split the velocity vector into a normal and a tangential part with respect
to the front; eq. (\ref{cons2}) then reads
    \begin{gather}
      \bar\rho\bar v_n=\alpha \rho_u v_{n,u}+(1-\alpha)\rho_b v_{n,b}
      \quad \makebox[0pt][l]{and}\\
      v_{t,u}=v_{t,b}=\bar{v}_t.
    \end{gather}

Further, the reconstructed states must satisfy the Rayleigh criterion and
the Hugoniot jump condition for the internal energy:
    \begin{gather}
      (\rho_u s_u)^2 = -\frac{p_b-p_u}{V_b-V_u} \\
      e_{i,b}-e_{i,u} =q -\frac{(p_b+p_u)}{2}(V_b-V_u)
    \end{gather}

Here, $e_i$ is defined as $e_{\text{tot}}-\vec v^2/2$ and $V:=1/\rho$.
The pressures are given by the equation of state:
    \begin{alignat}{4}
      \label{pcalc}
      p_u&=p_{\text{EOS}}(\rho_u&&, e_{i,u}&&, \mathbf{X}_u&&)
         \quad \text{and}\\
      p_b&=p_{\text{EOS}}(\rho_b&&, e_{i,b}&&, \mathbf{X}_b&&)
    \end{alignat}

Additionally, the jump condition for the normal velocity component reads
    \begin{equation}
      v_{n,b}-v_{n,u}=s_u\left(1-\frac{\rho_u}{\rho_b}\right).
    \end{equation}

To complete the system, a burning rate law is required. Usually
this will be the equation for the laminar burning speed, depending on the
unburnt state variables. In our case of highly turbulent burning in the
flamelet regime, the flame speed can be derived from the turbulent
kinetic sub-grid energy $e_{\text{sg}}$ \citep{niemeyer-hillebrandt-95a,
niemeyer-hillebrandt-95b}:
    \begin{equation}
      s_u= \sqrt{2e_{\text{sg}}}
    \end{equation}

The ensemble of all the equations above can be solved with any of standard
iterative method. Our implementation uses a globally converging Broyden
solver \citep{broyden-65, press-etal-92}. In contrast to the popular
Newton-Raphson approach, this algorithm converges even for relatively bad
initial guesses, which is important for our application.

\subsubsection{Transport}
The algorithms presented in the three following subsections are designed
for use with a \emph{directional splitting} scheme and are thus orientation
independent. Therefore we will only describe the numerical procedure
for the $x$-sweeps.

For the complete implementation a simple, non-con\-ser\-va\-tive approach is
used to obtain the $G$-values at the new time level:
    \begin{equation}
      \label{levupd}
      G^{n+1} = G^n -\Delta t D^n_x\frac{\partial G^n}{\partial x}
    \end{equation}
Several complications arise from the fact that values for $D_x$, which is
defined in the center of the mixed cells, are needed at the cell corners.
Since $\vec D$ only has a physical meaning in the mixed cells, its value
in all other cells may be chosen arbitrarily. It can be shown analytically that
the distance function property of $G$ is preserved if the
condition
    \begin{equation}
      \label{spread}
      \vec n \vec\nabla (\vec D \vec n)= 0
    \end{equation}
is satisfied, i.e. if the flame propagation velocity is constant along the
``field lines'' of $G$. Consequently, the values for $\vec D$ in the whole
computational domain are obtained by spreading out the values in the mixed
cells in the direction of $\vec n$ and $-\vec n$.

In the next step, $D_x$ in the middle of the cell interfaces is calculated
by simple averaging
    \begin{equation}
      D_{x,i,j+1/2}=\frac{1}{2} (D_{x,i,j}+D_{x,i,j+1}),
    \end{equation}
and the corner values $D_{x,i+1/2,j+1/2}$ are determined by upwinding.
Depending on the sign of $D_x$ at the corner, the appropriate discrete
derivative of $G$ is chosen; if $D_x$ is negative, one takes
$(\partial G/\partial x)$ at the right side, and vice versa.
Now all quantities needed in eq. (\ref{levupd}) are known.

Because of the discrete nature of the grid, it is in most cases impossible
to satisfy condition (\ref{spread}) exactly; therefore a re-initialization
step is required for the complete implementation also. This is done in
exactly the same fashion as described in section \ref{passive-reinit}.

\subsubsection{Flux-Splitting}
  \begin{figure}
  \centerline{\epsfig{file=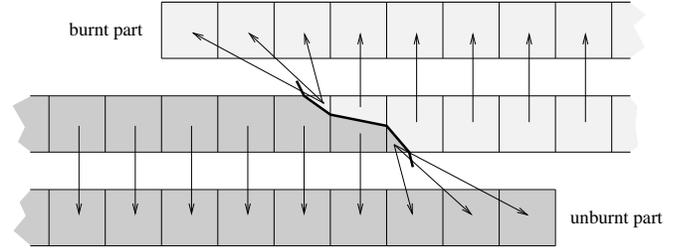,width=0.48\textwidth}}
  \caption{Splitting of a state vector containing burnt and unburnt
    cells into partial vectors with only fuel or ashes. The necessary
    ghost cells at the artificial boundaries are calculated by
    zeroeth order extrapolation.}
  \label{split}
  \end{figure}
In order to compute the total fluxes across a mixed cell interface, it is
necessary to solve the Riemann problems for burnt and unburnt states
separately. To achieve this, each grid vector is splitted into a sequence
of completely burnt and unburnt partial vectors. During this process,
artificial boundaries are created at the front location for which boundary
conditions must be specified. Following \cite{smiljanovski-etal-97}, this
is done by zeroeth order extrapolation of the cells at the boundary
(see Fig. \ref{split} for illustration). The PPM algorithm implemented in
PROMETHEUS is then used to calculate the hydrodynamical fluxes for the
partial vectors.

  \begin{figure}
  \centerline{\epsfig{file=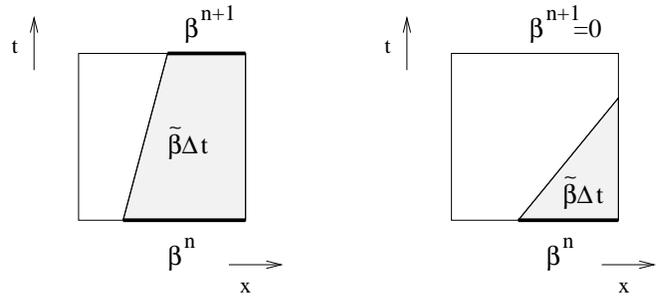,width=0.48\textwidth}}
  \caption{Determination of the average unburnt interface area fraction
   $\tilde\beta$ for two different cases. As can be seen, simply taking
   the average of old and new time level does not always produce the correct
   result.}
  \label{beta}
  \end{figure}
Now eq. (\ref{spliteq}) is applied to compose the total fluxes. However,
it is in many cases insufficient to use the unburnt interface fraction
$\beta$ at the beginning of the time step in this formula, especially
when the flame enters or leaves a cell during the time step.
Therefore we calculate the average of $\beta$ over the time step
(see also Fig. \ref{beta}):
\begin{equation}
\tilde\beta = \frac{1}{\Delta t} \int_{t^n}^{t^{n+1}}\beta dt
\end{equation}
The composed flux then reads
    \begin{equation}
      \bar\mathbf{F}=\tilde\beta \mathbf{F}_u+(1-\tilde\beta) \mathbf{F}_b.
    \end{equation}

\subsubsection{Source terms}
The amount of matter consumed by the flame in a mixed cell during a time step
is given by
      \begin{equation}
        \Delta m = \int_{t^n}^{t^{n+1}} A s_u \rho_u dt,
      \end{equation}
where $A$ denotes the flame surface in this cell. For the $x$-sweep in a
directional splitting scheme one obtains
      \begin{equation}
        \Delta m_x = \int_{t^n}^{t^{n+1}} n_x^2 A s_u \rho_u dt.
      \end{equation}

The factor $n_x^2$ is introduced by the projection of the flame on the
$y$-axis (or on the $yz$-plane in three dimensions) and by the projection
of the burning speed on the $x$-axis. In the $i$-th cell, the ratio of
the projected flame surface and the surface of a cell interface is
approximately given by $|\tilde\beta_{i+1/2}-\tilde\beta_{i-1/2}|$.
Thus one obtains for the source terms
      \begin{align}
        \Delta X_{\text{Ashes},i}&=\frac{\Delta t}{\Delta x_i}
           \frac{\rho_{u,i}}{\bar \rho_i}
          \left| s_{u,i}n_{x,i}(\tilde\beta_{i+1/2}-\tilde\beta_{i-1/2})
          \right|\\
        \Delta X_{\text{Fuel},i}&=-\Delta X_{\text{Ashes},i}\\
        \Delta e_{\text{tot},i}&=q \Delta X_{\text{Ashes},i}.
      \end{align}

\section{Numerical tests}
\label{test}
A set of testcases was run with both of the implementations presented above
to determine the ability of the numerical schemes to represent thermonuclear
flames. Our main criteria were the reproduction of a given burning velocity
and the isotropy of the front propagation. Additionally, we investigated
the behaviour of the algorithms for complex situations, like the
merging of two flame kernels and cusp formation in a sinusoidally perturbed
flame.

At $t=0$, the thermodynamical state of the unburnt matter was characterized
by $\rho_u=5\cdot 10^8$g/cm$^3$, $T_u=5\cdot 10^8$K, and $X_{^{12}C,u}=
X_{^{16}O,u}=0.5$. The energy release for the fusion to $^{56}$Ni is
$q=7\cdot 10^{17}$erg/g, and the burning speed $s_u$ was set to
$3\cdot 10^7$cm/s.

For all tests we used a cartesian grid with a cell size of $1.5\cdot 10^6$cm.

\subsection{1D flames}
  \begin{figure}
  \centerline{\epsfig{file=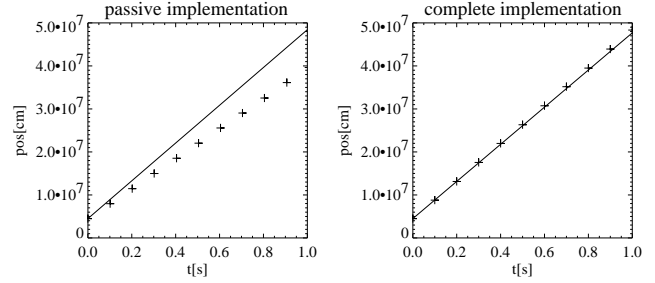,width=0.48\textwidth}}
  \caption{Time dependent position of a planar flame propagating in positive
    $x$-direction for both implementations of the front tracking algorithm.
    The lines indicate the theoretically predicted behaviour.}
  \label{levpos}
  \end{figure}

  \begin{figure*}[p]
  \centerline{\epsfig{file=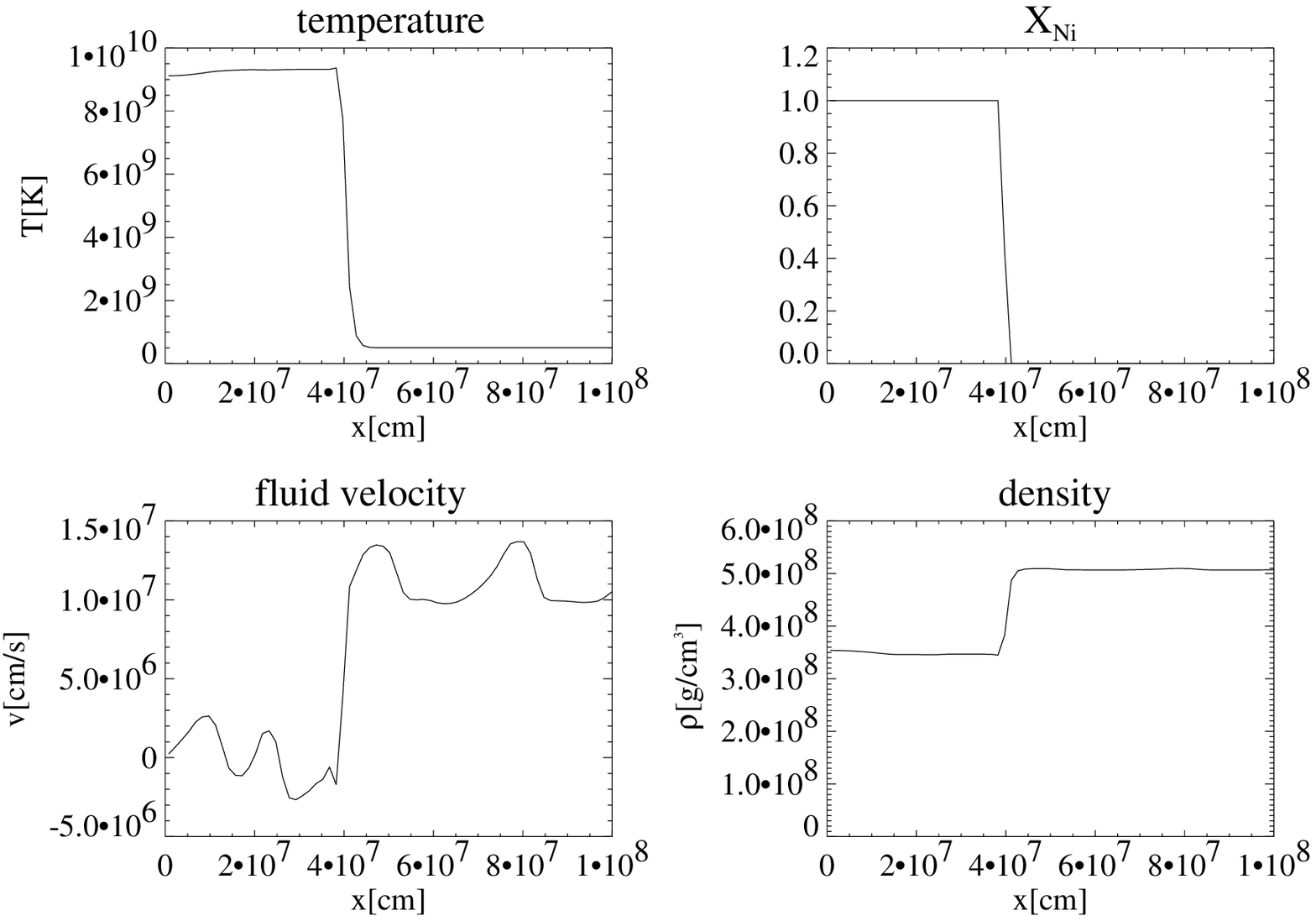,width=0.8\textwidth}}
  \caption{
    Planar flame propagation test: Temperature, nickel concentration,
    fluid velocity and density at $t=$1s for the passive implementation.}
  \label{planar-passive}
  \end{figure*}

  \begin{figure*}[p]
  \centerline{\epsfig{file=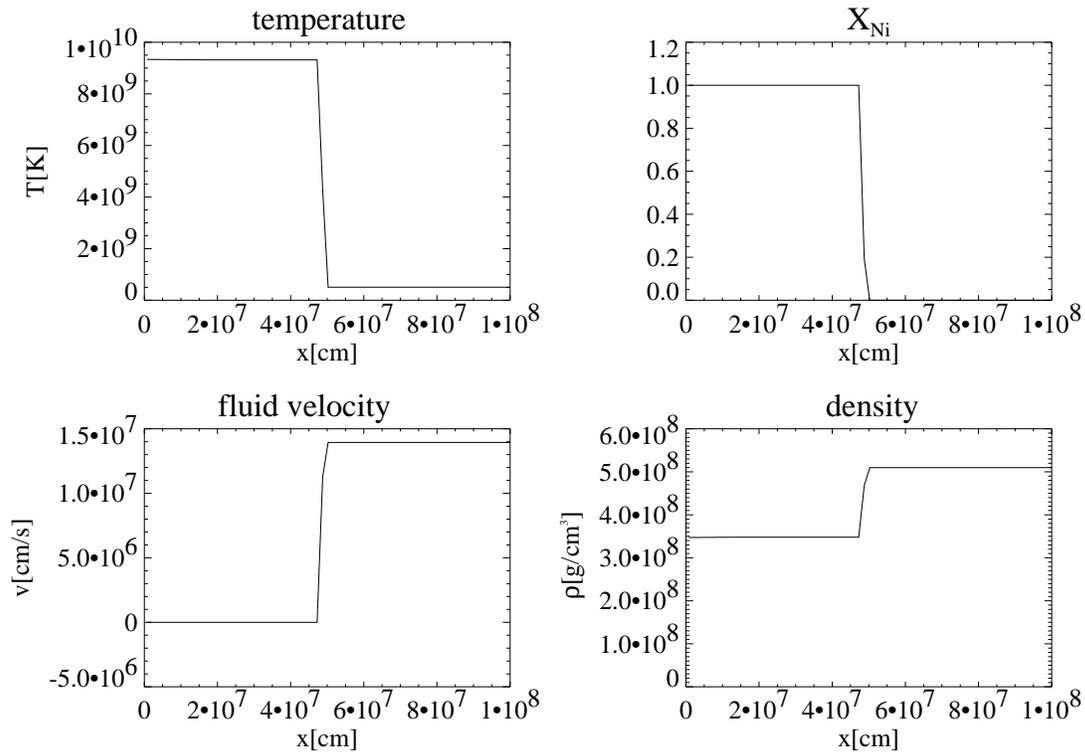,width=0.8\textwidth}}
  \caption{
    Planar flame propagation test: Temperature, nickel concentration,
    fluid velocity and density at $t=$1s for the complete implementation.}
  \label{planar-complete}
  \end{figure*}

In a first test we investigated a planar flame propagating in positive
$x$-direction with reflecting boundaries at the left, top and bottom edges
of the computational domain and an outflow boundary to the right.
The grid consisted of 128x4 cells.
Under these circumstances the material behind the
front should be at rest and the absolute front velocity with respect to
the grid is expected to be
\begin{equation}
s_b := s_u \rho_u/\rho_b,
\end{equation}
which corresponds to about $4.4\cdot 10^7$cm/s for our initial conditions.

As can be seen in Fig. \ref{levpos}, the agreement of simulation and
predictions is excellent for the complete implementation, whereas the passive
implementation underestimates the flame velocity by about 20\%. Figs.
\ref{planar-passive} and \ref{planar-complete} show the profiles of
temperature, nickel concentration, velocity and density for both algorithms
at $t=1$s. Again, the complete implementation gives exactly the expected
results: two constant states that are separated by a mixed cell. With the
exception of the fluid velocity, the picture is nearly the same for the passive
implementation; here the transition is smeared out over about three grid
cells by PPM. The velocity profile shows
strong oscillations in this case; one also notes that the average fluid
motion in the unburnt material is noticeably slower than for the complete
implementation.

All the deviations in the passive approach can be explained by the fact that
the flame is not advected with the speed of the unburnt matter as postulated
in eq. (\ref{Df}), but by the \emph{average} speed in the burning cells.
Depending on whether the flame just entered the cell or is about to leave it,
this quantity is closer to the unburnt resp. the burnt velocity but never
reaches the desired $\vec v_u$. As a consequence, the flame propagates
too slowly and at a non-uniform speed, thereby causing fluctuations in the
velocity field.

To further investigate this behaviour of the passive implementation, two
additional tests with $\rho_u=3\cdot 10^9$g/cm$^3$ and
$\rho_u=5\cdot 10^7$g/cm$^3$ were performed. In these cases the flame
propagation speed was underestimated by 14\% and 28\%, respectively. Since
the error grows roughly proportionally with the density jump, these
observations support our interpretation.

\pagebreak
\clearpage

However, for the special case of turbulent burning in the interior of white
dwarfs, these seemingly large errors can be tolerated:
firstly, the velocity jump across the flame is quite small compared to the
burning velocity; secondly, our model for the turbulent burning speed is based
on dimensional analysis and therefore $s_u$ itself could carry an uncertainty
much larger than the 28\% mentioned above.

A first order correction for the underestimation of the burning speed
in the passive implementation can be done in a quite straightforward way
for this concrete physical problem and will be incorporated in future
versions of the code.

\subsection{2D flames}

\subsubsection{One circular flame}
  \begin{figure}
  \centerline{\epsfig{file=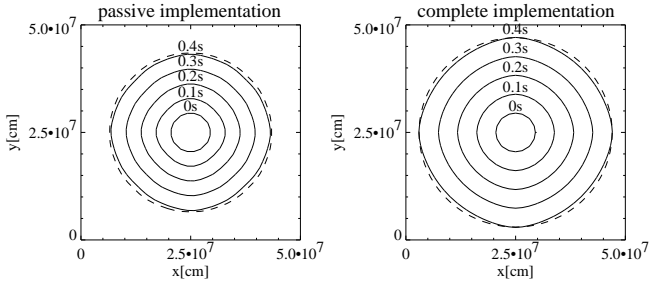,width=0.48\textwidth}}
  \caption{Snapshots of the front geometry for circular flame propagating
    outwards. The dashed lines represent
    exact circles and have been added to allow easier judgement of the flame
    geometry.}
  \label{circle}
  \end{figure}
To test the isotropy of both algorithms, the propagation of an
initially circular flame was simulated on a grid of 50x50 cells with
outflow boundaries; some snapshots of the front geometry are shown in
Fig. \ref{circle}.
While deviations from the circle shape do occur, they are sufficiently
small for both implementations. The difference in the flame propagation
speed is still present and nearly of the same size as in the one-dimensional
simulation.

\subsubsection {Two merging circular flames}
  \begin{figure}
  \centerline{\epsfig{file=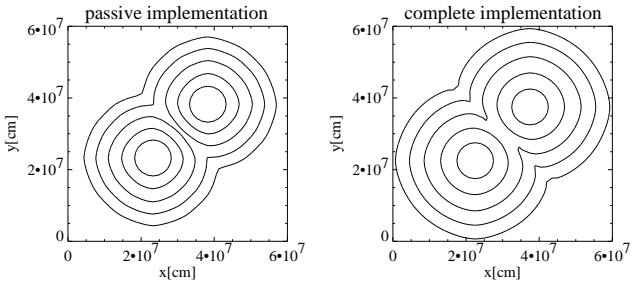,width=0.48\textwidth}}
  \caption{Evolution of the front geometry for two merging circular flames.
    The time difference
    between subsequent snapshots is 0.1s (from inside to outside).}
  \label{merging}
  \end{figure}
On the same grid as in the simulation above, the merging of two circular
flame kernels was investigated to demonstrate the ability of the level set
approach to handle topological changes. As the results indicate, the formation
of a single front happens smoothly and without numerical difficulties (see
Fig. \ref{merging}). The slight deformation of the fronts before the merging
can be explained by the interaction of the velocity fields generated by both
flames.

\subsubsection {Perturbed planar flame}
  \begin{figure}
  \centerline{\epsfig{file=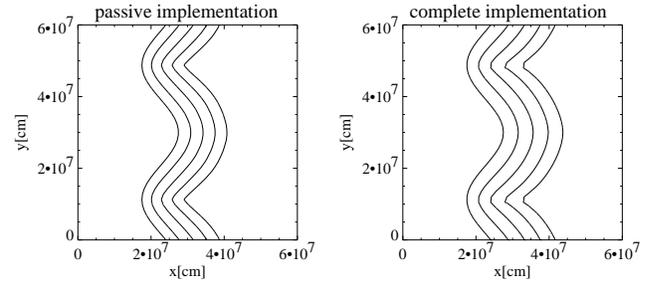,width=0.48\textwidth}}
  \caption{Evolution of the front geometry for a sinusoidally perturbed
    flame. The time
    difference between subsequent snapshots is 0.1s (from left to right).}
  \label{cusp}
  \end{figure}
Fig. \ref{cusp} shows the temporal evolution of a sinusoidally perturbed
flame propagating in positive $x$-direction. As expected, the trailing part
of the front becomes narrower until a cusp is formed; afterwards, the flame
geometry remains practically unchanged. The short vertical section of the
flame that can be seen in the right panel of Fig. \ref{cusp} is an artifact
of the rather poor resolution: since the (expected) cusps are located
exactly at the $y$-position of the cell centers and the level set
is stored at the cell corners, they cannot be seen in this discretization.

\subsection{Sensitivity of the reconstruction equations}
The results of all tests described above show that both implementations
of the level set method can be used to model turbulent thermonuclear
combustion in Type Ia supernovae. Since the complete version is more
accurate, it would be the method of choice.
Unfortunately, however, it has turned out that the straightforward
implementation of the reconstruction as described above leads to numerical
difficulties when
applied to the ``real'' situation in a white dwarf including density and
pressure gradients and gravitational forces. In our supernova simulations
with the complete implementation, after a few time steps the reconstruction
in many mixed cells failed because the internal energy of the unburnt state
reached values for which the equation of state is undefined:
\begin{equation}
e_{i,u} < e_{\text{EOS}}(\rho_u, T=0\text{K}, X_{\text{fuel}})
\end{equation}

\begin{figure}
  \centerline{\epsfig{file=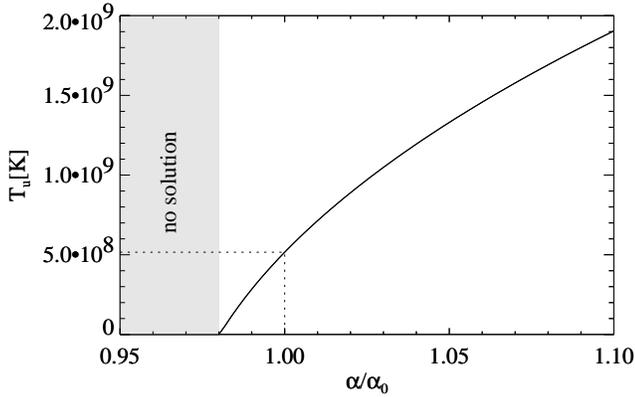,width=0.48\textwidth}}
  \caption{Reconstructed temperature of the unburnt material for varying
    deviations in $\alpha$. Below $\alpha/\alpha_0 < 0.98$, the reconstruction
    fails.}
  \label{alphatest}
  \end{figure}

Discretization errors in the input values are the most likely reason for
this divergence. To test the reaction of the reconstruction algorithm on
such uncertainties, the following experiment was performed:

From a given pre- and post-front state that exactly fulfill the
Rankine-Hugoniot jump conditions a mixed state is synthesized according
to eqs. (\ref{cons1})-(\ref{cons3}) for an $\alpha_0$ equal to 0.5.
Then a reconstruction is tried for this mixed state, but for a slightly
different $\alpha$ (i.e. for an $\alpha$ with some uncertainty).
In Fig. \ref{alphatest}, the reconstructed temperature of the unburnt material
is plotted against the introduced error in $\alpha$. It can be easily seen
that for $\alpha/\alpha_0 < 0.98$ a reconstruction of pre- and
post-front states becomes impossible. For highly curved fronts, as they are
expected in Type Ia supernovae, the deviations of $\alpha$ from the exact value
can become much higher than that, because $\alpha$ is obtained for a
piecewise linear approximation of the front.
At first glance, one would expect an improvement if the front geometry
was modeled with higher accuracy, e.g. by approximation with quadrics.
But in this case, other problems appear: the number of different topological
configurations in a cell explodes, and, most importantly, there is no way to
define the normal $\vec n$ that is required by the reconstruction equations.

Because of these numerical problems, we have not yet been able to simulate
Type Ia supernovae with the complete implementation of the front tracking
scheme. An investigation of the properties of the reconstruction equations and,
if possible, creation of a more robust system is subject of future work.
However, introducing just an artificial viscosity to limit the curvature of
the flame front may be an easy way to stabilize the numerical scheme.

\section{Applications}
\label{appl}
\subsection{Type Ia supernovae}

\begin{figure*}[p]
  \centerline{\epsfig{file=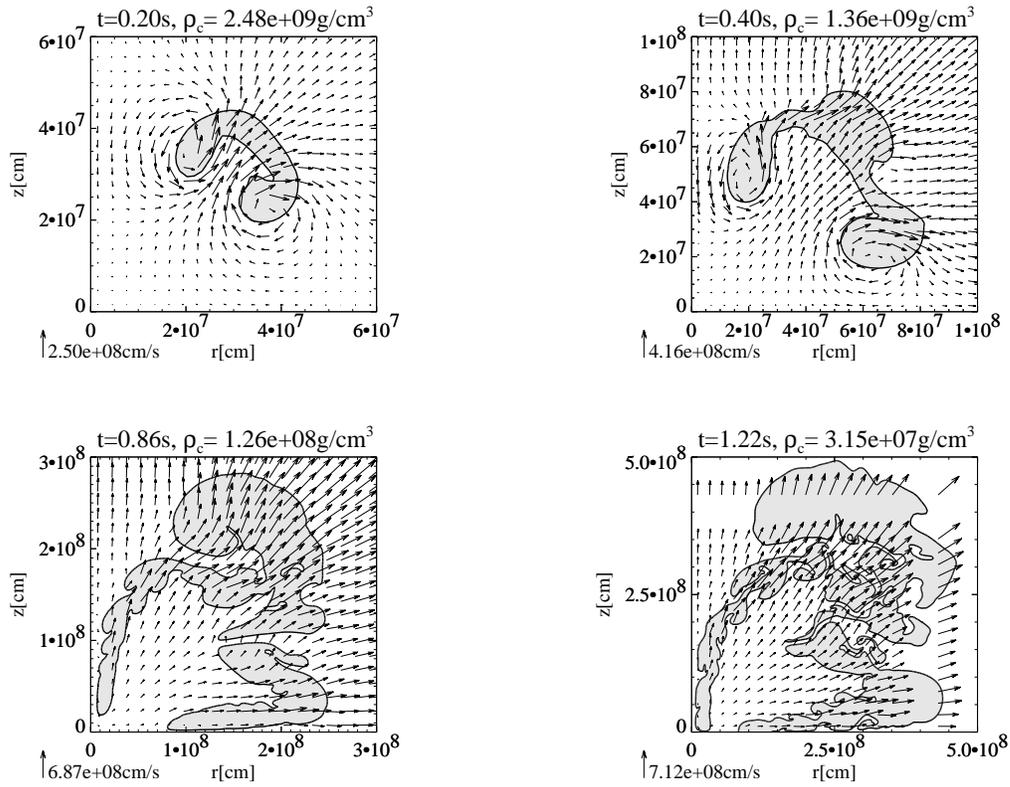, height=0.45\textheight}
  \hspace*{1cm}}
  \caption{Temporal evolution of front geometry and velocity field after
    igniting a single, circular bubble near the center of the white dwarf.
    Note that the scales change from snapshot to snapshot.}
  \label{b1fr}
  \end{figure*}

The passive implementation of the level set method has been used to model
the turbulent flame front in the early stages of Type Ia supernova explosions.
To allow direct comparison with the reaction-diffusion model, our initial
conditions were chosen as similar as possible to the simulations done by
\cite{niemeyer-hillebrandt-95b}. Our results show a flame which is perturbed
due to Rayleigh-Taylor- and Kelvin-Helmholtz-instabilities on all scales down
to a few grid cells (see Fig. \ref{b1fr}). An extensive discussion of this
simulation as well as simulations with other initial conditions can be found
in \cite{reinecke-etal-98b}.

\subsection{Chemical hydrogen combustion}
\begin{figure*}[p]
  \centerline{\epsfig{file=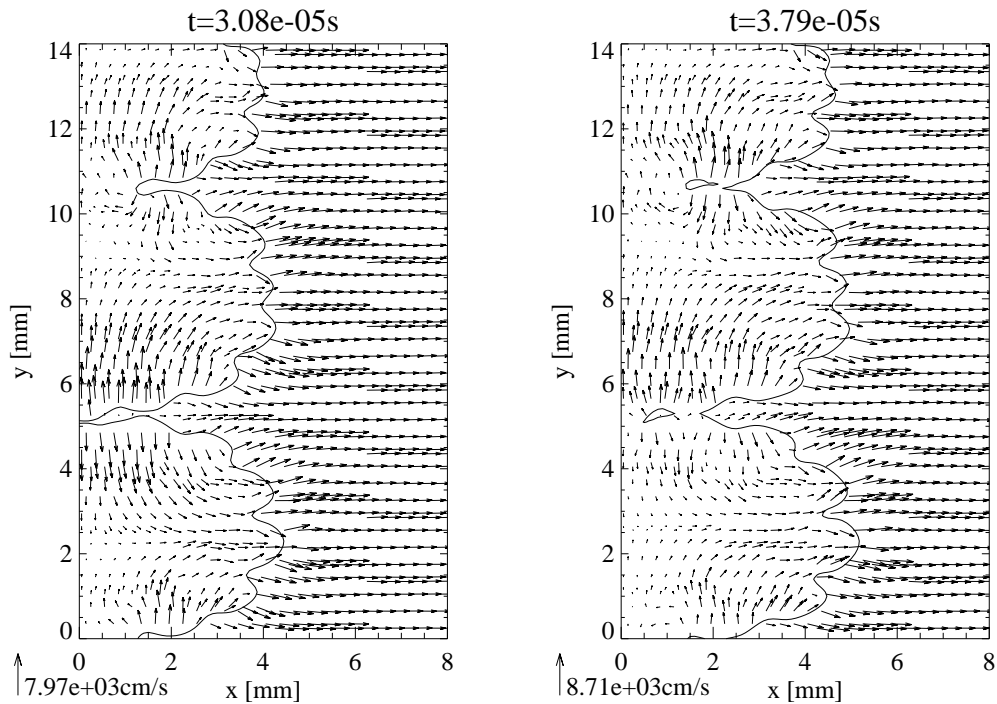, width=0.8\textwidth}
   \hspace*{1cm}}
  \caption{Chemical combustion in a lean hydrogen-air mixture on a grid
    consisting of 100$\times$140 cells, with reflecting borders at
    the left, top and bottom and a flowout boundary at the right.
    Due to the high material diffusion of hydrogen, little disturbances
    are amplified. This was modeled by using a curvature-dependent flame speed.}
  \label{h2}
  \end{figure*}

The complete implementation has already been successfully used to model
turbulent flame fronts in lean hydrogen-air mixtures. Fig. \ref{h2} shows the
merging of three flame parts in a mixture of 15\% hydrogen in air in a box with
an
outflow boundary to the right and reflecting boundaries elsewhere. Since small
disturbances are amplified by material diffusion in the case of hydrogen flames,
the burning speed was modified depending on the curvature of the front.

\section{Conclusions}
\label{summ}
We have presented a numerical model to describe deflagration fronts
with a reaction zone much thinner than the cells of the computational grid.
In contrast to the currently favoured method for astrophysical simulations
\citep{khokhlov-93}, our approach provides a considerably sharper transition
from fuel to ashes, thereby allowing the growth of hydrodynamical instabilities
on smaller scales and generally the evolution of small features in the flame.

Two different implementations of the model have been developed and tested;
for simple initial conditions, both versions produce results acceptable for
our needs. However, because of the mentioned numerical problems the complete
implementation cannot be employed for supernova simulations without
modification.

In addition to modeling flames, the level set method described in this paper
can also be used for tracking contact discontinuities with only
minor modifications; therefore, any application in astrophysical hydrodynamics
dealing with one of these phenomena might benefit from this numerical scheme.

\begin{acknowledgements}
This work was supported in part by the Deutsche Forschungsgemeinschaft under
Grant Hi 534/3-1 and by DOE under contract No. B341495 at the University of
Chicago. The computations were performed at the Rechenzentrum Garching
on a Cray J90.

The authors thank E.\ Bravo for many constructive suggestions which led
to a significant improvement of this paper.
\end{acknowledgements}

\clearpage

\bibliographystyle{aabib}
\bibliography{refs}

\begin{thebibliography}{16}
\expandafter\ifx\csname natexlab\endcsname\relax\def\natexlab#1{#1}\fi

\bibitem[Broyden(1965)]{broyden-65}
Broyden, C.~G.: 1965, Mathematics of Computation 19, 577

\bibitem[Fryxell et~al.(1989)Fryxell, M\"uller \& Arnett]{fryxell-etal-89}
Fryxell, B.~A., M\"uller, E., Arnett, W.~D.: 1989, MPA Preprint 449

\bibitem[Khokhlov(1993)]{khokhlov-93}
Khokhlov, A.~M.: 1993, ApJ 419, L77

\bibitem[Mulder et~al.(1992)Mulder, Osher \& Sethian]{mulder-etal-92}
Mulder, W., Osher, S., Sethian, J.~A.: 1992, JCP 100, 209

\bibitem[Niemeyer(1994)]{niemeyer-94}
Niemeyer, J.~C.: 1994, Turbulente thermonukleare Brennfronten in Wei{\ss}en
  Zwergen, Master's thesis, Max-Planck-Institut f\"ur Astrophysik, Garching

\bibitem[Niemeyer(1995)]{niemeyer-95}
Niemeyer, J.~C.: 1995, On the Propagation of Thermonuclear Flames in Type Ia
  Supernovae, Ph.D. thesis, Max-Planck-Institut f\"ur Astrophysik, Garching

\bibitem[Niemeyer \& Hillebrandt(1995{\natexlab{a}})]{niemeyer-hillebrandt-95a}
Niemeyer, J.~C., Hillebrandt, W.: 1995{\natexlab{a}}, ApJ 452, 769

\bibitem[Niemeyer \& Hillebrandt(1995{\natexlab{b}})]{niemeyer-hillebrandt-95b}
Niemeyer, J.~C., Hillebrandt, W.: 1995{\natexlab{b}}, ApJ 452, 779

\bibitem[Nomoto et~al.(1984)Nomoto, Thielemann \& Yokoi]{nomoto-etal-84}
Nomoto, K., Thielemann, F.~K., Yokoi, K.: 1984, ApJ 286, 644

\bibitem[Osher \& Sethian(1988)]{osher-sethian-88}
Osher, S., Sethian, J.~A.: 1988, JCP 79, 12

\bibitem[Press et~al.(1992)Press, Teukolsky, Vetterling \&
  Flannery]{press-etal-92}
Press, W.~H., Teukolsky, S.~A., Vetterling, W.~T., Flannery, B.~P.: 1992,
  Numerical Recipes in C, Cambridge: Cambridge University Press

\bibitem[Reinecke et~al.(1998)Reinecke, Hillebrandt \&
  Niemeyer]{reinecke-etal-98b}
Reinecke, M.~A., Hillebrandt, W., Niemeyer, J.~C.: 1998, MPA Preprint 1122b,
  accepted by A\&A

\bibitem[Sethian(1996)]{sethian-96}
Sethian, J.~A.: 1996, Level Set Methods, Cambridge: Cambridge University Press

\bibitem[Smiljanovski et~al.(1997)Smiljanovski, Moser \&
  Klein]{smiljanovski-etal-97}
Smiljanovski, V., Moser, V., Klein, R.: 1997, Comb. Theory and Modeling 1, 183

\bibitem[Sussman et~al.(1994)Sussman, Smereka \& Osher]{sussman-etal-94}
Sussman, M., Smereka, P., Osher, S.: 1994, JCP 114, 146

\bibitem[Timmes \& Woosley(1992)]{timmes-woosley-92}
Timmes, F.~X., Woosley, S.~E.: 1992, ApJ 396, 649

\end{thebibliography}

\end{document}